\documentclass[preprint,superscriptaddress,preprintnumbers,showpacs]{revtex4}
\usepackage{amssymb}
\usepackage{graphicx}
\usepackage{dcolumn}
\usepackage{bm}

\begin{document}

\newcommand*{\sjtu}{INPAC, Department of Physics and Shanghai Key Laboratory for Particle Physics and Cosmology, Shanghai Jiao Tong University,  Shanghai, China}\affiliation{\sjtu}
\newcommand*{\NTU}{CTS, CASTS and Department of Physics, National Taiwan University, Taipei, Taiwan}\affiliation{\NTU}
\newcommand*{\NCTS}{Department of Physics, National Tsing Hua University, and National Center for Theoretical Sciences, Hsinchu, 300, Taiwan}\affiliation{\NCTS}

\title{The $\alpha$, $\beta$ and $\gamma$ parameterizations of  CP violating CKM phase}

\author{Guan-Nan Li}\affiliation{\sjtu}
\author{Hsiu-Hsien Lin}\affiliation{\NTU}
\author{Dong Xu}\affiliation{\sjtu}
\author{Xiao-Gang He}\email{hexg@phys.ntu.edu.tw}\affiliation{\sjtu}\affiliation{\NTU}\affiliation{\NCTS}

\begin{abstract}
The CKM matrix describing quark mixing with three generations can be
parameterized by three mixing angles and one CP violating phase. In
most of the parameterizations, the CP violating phase chosen is not
a directly measurable quantity and is parametrization dependent. In
this work, we propose to use experimentally measurable CP violating
quantities, $\alpha$, $\beta$ or $\gamma$ in the unitarity triangle
as the phase in the CKM matrix, and  construct explicit $\alpha$,
$\beta$ and $\gamma$ parameterizations. Approximate Wolfenstein-like
expressions are also suggested.

\end{abstract}


\maketitle

\section{Introduction}
The mixing between different quarks is
described by an unitary matrix in the charged current
interaction of W-boson in the mass eigen-state of quarks,  the
Cabibbo~\cite{cabibbo}-Kobayashi-Maskawa~\cite{km}(CKM) matrix
$V_{\rm{CKM}}$, defined by
\begin{eqnarray}
L = -{g\over \sqrt{2}} \overline{U}_L \gamma^\mu V_{\rm CKM} D_L
W^+_\mu  + H.C.\;,
\end{eqnarray}
where $U_L = (u_L,c_L,t_L,...)^T$, $D_L = (d_L,s_L,b_L,...)^T$. For n-generations, $V =
V_{\rm CKM}$ is an $n\times n$ unitary matrix. With three generations, one can write
\begin{eqnarray}
V_{\rm CKM} = \left ( \begin{array}{lll}
V_{ud}&V_{us}&V_{ub}\\
V_{cd}&V_{cs}&V_{cb}\\
V_{td}&V_{ts}&V_{tb}
\end{array}
\right )\;.
\end{eqnarray}

A commonly used parametrization for mixing matrix with three generations of quark is given by~\cite{ck,yao},
\begin{eqnarray}
V_{PDG} = \left(
\begin{array}{ccc}
c_{12}c_{13} & s_{12}c_{13} & s_{13}e^{-i\delta_{PDG}}           \\
-s_{12}c_{23}-c_{12}s_{23}s_{13}e^{i\delta_{PDG}} &
c_{12}c_{23}-s_{12}s_{23}s_{13}e^{i\delta_{PDG}}  & s_{23}c_{13} \\
s_{12}s_{23}-c_{12}c_{23}s_{13}e^{i\delta_{PDG}}  &
-c_{12}s_{23}-s_{12}c_{23}s_{13}e^{i\delta_{PDG}} & c_{23}c_{13}
\end{array}
\right),\label{fp}
\end{eqnarray}
where $s_{ij}=\sin\theta_{ij}$ and $c_{ij}=\cos\theta_{ij}$ with $\theta_{ij}$ being angles rotating in flavor space
and $\delta_{PDG}$ is the CP violating phase. We refer this as the PDG parametrization.

There are a lot of experimental data on the mixing pattern of
quarks. Fitting available data, the mixing angles and CP violating phase are determined to be~\cite{utfit}
\begin{eqnarray}
&&\theta_{12}=13.015^\circ\pm0.059^\circ,\quad
\theta_{23}=2.376^\circ \pm0.046^\circ,\quad
\theta_{13}=0.207^\circ\pm0.008^\circ, \nonumber\\
&&\delta_{PDG}=69.7^\circ\pm3.1^\circ. \label{qangle}
\end{eqnarray}

From the above, we obtain the magnitude of the matrix elements as
\begin{eqnarray} \left(
  \begin{array}{lll}
    0.9743\pm 0.0002             & 0.2252\pm 0.0010    &0.0036\pm 0.0001              \\
    0.2251\pm 0.0010             & 0.9735 \pm 0.0002  & 0.0415\pm 0.0008      \\
    0.0088\pm 0.0003              & 0.0407\pm 0.0008   & 0.99913\pm 0.00003
  \end{array} \right)\;.\label{vv}
\end{eqnarray}

The angles can be viewed as rotations in flavor spaces. But both the
angles and the phase in the CKM matrix are not directly measurable
quantities. There are different ways to parameterize the mixing
matrix. In different parametrizations, the angles and phase are
different. To illustrate this point let us study the original KM
parametrization~\cite{km},
\begin{eqnarray}
V_{KM} = \left ( \begin{array}{ccc} c_1& - s_1 c_3& -s_1 s_3\\s_1c_2&c_1c_2c_3 - s_2s_3 e^{i\delta_{KM}}&c_1c_2s_3 + s_2c_3 e^{i\delta_{KM}}\\
s_1s_2&c_1s_2c_3 + c_2 s_3 e^{i\delta_{KM}}& c_1s_2 s_3 - c_2c_3 e^{i\delta_{KM}}\end{array}
\right )\;.
\end{eqnarray}
Using the observed values for the mixing matrix, one would obtain
\begin{eqnarray}
&&\theta_1 = 13.016^\circ\pm0.003^\circ\;,\;\;\theta_2 = 2.229^\circ\pm0.066^\circ\;,\;\;\theta_3 =0.921^\circ\pm0.036^\circ\;,
\end{eqnarray}
and the central value of the CP violating phase angle is $\delta_{KM} = 88.2^\circ$.

\begin{figure}[htbp]
\centering
\includegraphics[width = 0.6\textwidth]{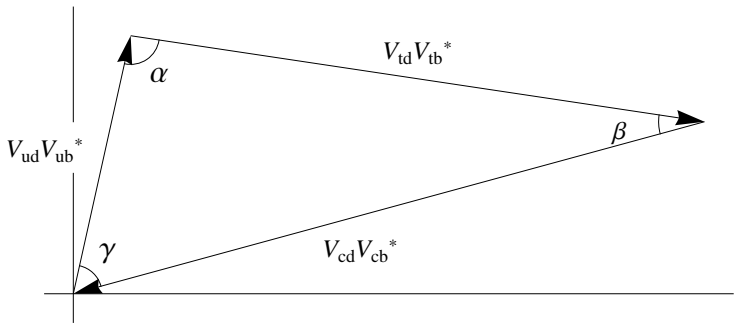}\hspace{0.5cm}
\caption{The unitarity triangle.}\label{fig}
\end{figure}

We see that the angles and phases in the PDG and KM
parameterizations are indeed very different. The angles and phase
are parametrization dependent. It is interesting to see whether all
quantities used to parameterize the mixing matrix can all have well
defined physical meanings, that is, all are experimentally
measurable quantities, as have been done for several other
quantities related to mixing
matrices~\cite{jarlskog,triangle,fh,qlc}. To this end we notice that
the magnitudes of the CKM matrix elements are already experimentally
measurable quantities, one can take them to parameterize the mixing
matrix. However, the information on CP violation is then hid in the
matrix elements, a single magnitude of an element would not be able
to be taken as the measure of CP violation. Only a combination of
several magnitude is able to signify the CP violation. For example,
$Am = 1- (|V_{td}|^2|V_{tb}|^2 + |V_{ud}|^2|V_{ub}|^2 -
|V_{cd}|^2|V_{cb}|^2)^2/4 |V_{td}|^2|V_{tb}|^2|V_{ud}|^2|V_{ub}|^2$
is non-zero. Or one needs to know the phase of several matrix
elements such as $J\sum_{n,m = 1}^3
\epsilon_{ikm}\epsilon_{jln}=Im(V_{ij}V_{kl}V^*_{il}V^*_{kj})$
\cite{jarlskog}~is non-zero. Experimentally there are several
measurable phases which can signify CP violations. The famous ones
are the angles $\alpha$, $\beta$ and $\gamma$ in the unitarity
triangle defined by the unitarity condition
\begin{eqnarray}
V_{ud}V_{ub}^* + V_{cd}V_{cb}^* + V_{td}V^*_{tb}=0
\end{eqnarray}
In the complex plane, the above defines a triangle shown in Fig 1.

This triangle defines three phase angles
\begin{eqnarray}
\ {\alpha} = {\arg\left
(-{{V_{td}V_{tb}^*}\over{V_{ud}V_{ub}^*}}\right )}\;,\;\;
\ {\beta} = {\arg\left
(-{{V_{cd}V_{cb}^*}\over{V_{td}V_{tb}^*}}\right )}\;,\;\;
\ {\gamma} = {\arg\left
(-{{V_{ud}V_{ub}^*}\over{V_{cd}V_{cb}^*}}\right )}\;. \label{phases}
\end{eqnarray}

CP violation dictates that the area of this triangle to be non-zero. This
implies that none of the angles $\alpha$, $\beta$ and $\gamma$ can be zero.
Experimentally these three angles have been measured directly~\cite{yao},
$\alpha =(89.0^{+4.4}_{-4.2})^{\circ} $, $\beta =(21.1\pm0.9)^{\circ} $ and $\gamma =(73^{+22}_{-25})^{\circ} $.
These numbers are consistent with that obtained using the numerical numbers in eq. \ref{qangle}, $\alpha ={88.14}^{\circ} $, $\beta ={22.20}^{\circ} $ and $\gamma ={69.67}^{\circ} $.
Also the directly measured numbers are consistent with the
SM prediction $\alpha + \beta +\gamma = \pi$.
Notice that the values $\alpha$, $\gamma$ are very close to the two phases
$\delta_{KM}$, $\delta_{PDG}$, respectively. We will see later that although they are
close to each other, they are  not exactly equal.

Since we know that one only needs one quantity to signify the existence of CP violation, $\alpha$, $\beta$ and $\gamma$ must be related. In fact they are related to the parameter $J$ as~\cite{triangle}
\begin{eqnarray}
J=|V_{td}||V_{tb}^*||V_{ud}||V_{ub}^*| sin\alpha = |V_{td}||V_{tb}^*||V_{cd}||V_{cb}^*| sin\beta = |V_{cd}||V_{cb}^*||V_{ud}||V_{ub}^*| sin\gamma\;.\label{relat1}
\end{eqnarray}
The above is twice of the triangle area in Fig 1.
The quantity $Am$ is equal to $\sin^2\alpha$.

To have a parametrization for $V_{CKM}$ in which all quantities are experimentally measurable ones, one can choose three
modulus of $|V_{ij}|$ and one of the above CP violating parameters. For the CP violating phase, it is clear that the
three phase angles $\alpha$, $\beta$ and $\gamma$ are among the simplest and have clear geometric meaning. We refer these
as the $\alpha$, $\beta$ and $\gamma$ parameterizations. In the following we discuss how
the parametrization can be constructed and how they can be transformed from each other.

\section{The $\alpha$, $\beta$ and $\gamma$ parametrizations}

In the $\alpha$ parametrization we take $\alpha$ as the phase appearing in the CKM matrix along with three modulus of $V_{ij}$.
From the definition of $\alpha$ in eq.\ref{phases}, we can have four ways in which only one of the $V_{ud, ub, td, tb}$ relevant to the definition of $\alpha$ is
complex and all others are real and positive,
\begin{eqnarray}
&&\alpha_1)\;. \;\;(|V_{ud}|, |V_{ub}|, |V_{td}|, -|V_{tb}|e^{-i\alpha})\;,\nonumber\\
&&\alpha_2)\;. \;\;(|V_{ud}|, |V_{ub}|, -|V_{td}|e^{i\alpha}, |V_{tb}|)\;,\nonumber\\
&&\alpha_3)\;. \;\;(|V_{ud}|, -|V_{ub}|e^{i\alpha}, |V_{td}|, V_{tb}|)\;,\nonumber\\
&&\alpha_4)\;. \;\;(- |V_{ud}|e^{-i\alpha}, |V_{ub}|, |V_{td}|, |V_{tb}|)\;.
\end{eqnarray}
In the above one can change the signs of the elements by defining quark phases.

Indicating the CKM matrix for the four cases by $V^{\alpha_i}_{CKM}$. We have
\begin{equation}
V^{\alpha_1}_{CKM} = \left(
\begin{array}{ccc}
|V_{ud}|& |V_{us}| & |V_{ub}|\\
|V_{cd}|&
 - \frac{(|V_{us}|^{2}-|V_{td}|^{2})|V_{ud}|+|V_{ub}||V_{td}||V_{tb}|e^{-i\alpha}}{|V_{us}||V_{cd}|}&\frac{|V_{td}||V_{tb}|e^{-i\alpha}-|V_{ud}||V_{ub}|}{|V_{cd}|}\\
 |V_{td}| & \frac{|V_{ub}||V_{tb}|e^{-i\alpha}-|V_{ud}||V_{td}|}{|V_{us}|}& - |V_{tb}|e^{-i\alpha}\\
\end{array}
\right)
\end{equation}
One can take $\alpha$,~$|V_{ud}|$,~$|V_{us}|$,~$|V_{cd}|$~as the four independent variables to parameterize the CKM matrix. The other elements
can be expressed as functions of them with
\begin{eqnarray}
|V_{td}|&=&\sqrt{1-|V_{ud}|^{2}-|V_{cd}|^{2}},\;\;|V_{ub}|=\sqrt{1-|V_{ud}|^{2}-|V_{us}|^{2}},\nonumber\\
|V_{tb}|&=&\frac{|V_{td}||V_{ud}||V_{ub}|\cos{\alpha}}{1-|V_{ud}|^{2}}\\
&+&
\sqrt{(\frac{|V_{td}||V_{ud}||V_{ub}|\cos{\alpha}}{1-|V_{ud}|^{2}})^{2}-\frac{|V_{cd}|^{2}(|V_{ub}|^{2}-1)+|V_{ud}|^{2}|V_{ub}|^{2}}{1-|V_{ud}|^{2}}}\;.\nonumber
\end{eqnarray}

For the other three cases, $V_{CKM}$ are given by
\begin{eqnarray}
&&V^{\alpha_2}_{CKM} = \left(
\begin{array}{ccc}
|V_{ud}|& |V_{us}| & |V_{ub}|\\
\frac{ |V_{td}||V_{tb}|e^{i\alpha}-|V_{ud}||V_{ub}|}{|V_{cb}|}&
 \frac{(|V_{ud}|^{2}-|V_{cb}|^{2})|V_{ub}|-|V_{ud}||V_{td}||V_{tb}|e^{i\alpha}}{|V_{us}||V_{cb}|}&|V_{cb}|\\
 -|V_{td}|e^{i\alpha} & \frac{|V_{ud}||V_{td}|e^{i\alpha}-|V_{ub}||V_{tb}|}{|V_{us}|} & |V_{tb}|\\
\end{array}
\right)\;,\nonumber\\
&&V^{\alpha_3}_{CKM} = \left(
\begin{array}{ccc}
|V_{ud}|&- \frac{|V_{ud}||V_{td}|-|V_{tb}||V_{ub}|e^{i\alpha}}{|V_{ts}|} & - |V_{ub}|e^{i\alpha}\\
|V_{cd}|&
 \frac{(|V_{tb}|^{2}-|V_{cd}|^{2})|V_{td}|-|V_{ub}||V_{ud}||V_{tb}|e^{i\alpha}}{|V_{ts}||V_{cd}|}&\frac{|V_{ud}||V_{ub}|e^{i\alpha}-|V_{td}||V_{tb}|}{|V_{cd}|}\\
 |V_{td}| & |V_{ts}|& |V_{tb}|\\
\end{array}
\right)\;,\\
&&V^{\alpha_4}_{CKM} = \left(
\begin{array}{ccc}
- |V_{ud}|e^{-i\alpha}& - \frac{|V_{ub}||V_{tb}|-|V_{td}||V_{ud}|e^{-i\alpha}}{|V_{ts}|} & |V_{ub}|\\
- \frac{|V_{td}||V_{tb}|-|V_{ud}||V_{ub}|e^{-i\alpha}}{|V_{cb}|}&
  -\frac{(|V_{cb}|^{2}-|V_{td}|^{2})|V_{tb}|+|V_{ud}||V_{td}||V_{ub}|e^{-i\alpha}}{|V_{ts}||V_{cb}|}&|V_{cb}|\\
 |V_{td}| & |V_{ts}| & |V_{tb}|\\
\end{array}
\right)\;.\nonumber
\end{eqnarray}

Similar to case $\alpha_1$, one can choose the phase $\alpha$ and three modulus of $V_{ij}$ as independent variables for the above three cases.
It is convenient to choose the parameter sets ($\alpha$,~$|V_{us}|$,~$|V_{ub}|$,~$|V_{cb}|$), ($\alpha$,~$|V_{ts}|$,~$|V_{cd}|$,~$|V_{td}|$)
and ($\alpha$,~$|V_{cb}|$,~$|V_{tb}|$,~$|V_{ts}|$) as independent variables for the above three cases, respectively.

In all the above four cases, the Jarlskog parameter $J$ is given by
\begin{eqnarray*}
J=|V_{ub}||V_{ud}||V_{td}||V_{tb}|\sin{\alpha}
\end{eqnarray*}
This is not surprising because the above four cases are equivalent.

To see the above four cases discussed are equivalent explicitly, let us demonstrate how one can transform case $\alpha_1$ to case $\alpha_2$ by redefining quark phases.
The different parameterizations are equivalent implies that by redefining quark phases, one can transform the different ways of parameterizations for $V_{CKM}^{\alpha_i}$ from one to another, that is,
\begin{eqnarray}
{V^{\alpha_{i}}_{CKM}}=
\left(
\begin{array}{ccc}
1 & 0 &0           \\
0 &e^{im} &0 \\
0&0 &e^{in}
\end{array}
\right)
{V^{\alpha_{j}}_{CKM}}
\left(
\begin{array}{ccc}
e^{ix} & 0 &0           \\
0 &e^{iy} &0 \\
0&0 &e^{iz}
\end{array}
\right),
\end{eqnarray}
where $i$ and $j$ stand for various types of parametrization.
For example, transforming case $\alpha_1$ to case $\alpha_2$ becomes a mission of finding the different parameters such that\\
\begin{eqnarray}
&&\quad\quad\quad\quad\quad\quad\quad\quad{V^{\alpha_{2}}_{CKM}}=
\left(
\begin{array}{ccc}
1 & 0 &0           \\
0 &e^{im} &0 \\
0&0 &e^{in}
\end{array}
\right)
{V^{\alpha_{1}}_{CKM}}
\left(
\begin{array}{ccc}
e^{ix} & 0 &0           \\
0 &e^{iy} &0 \\
0&0 &e^{iz}
\end{array}
\right)
\\
&=&\left(
\begin{array}{ccc}
|V_{ud}|e^{ix}& |V_{us}|e^{iy} & |V_{ub}|e^{iz}\\
|V_{cd}|e^{i(m+x)}&
 - \frac{(|V_{us}|^{2}-|V_{td}|^{2})|V_{ud}|e^{i(m+y)}+|V_{ub}||V_{td}||V_{tb}|e^{i(m+y-\alpha)}}{|V_{us}||V_{cd}|}&\frac{|V_{td}||V_{tb}|e^{i(m+z-\alpha)}-|V_{ud}||V_{ub}|e^{i(m+z)}}{|V_{cd}|}\\
 |V_{td}|e^{i(n+x)} & \frac{|V_{ub}||V_{tb}|e^{i(n+y-\alpha)}-|V_{ud}||V_{td}|e^{i(n+y)}}{|V_{us}|}& - |V_{tb}|e^{i(n+z-\alpha)}\\
\end{array}
\right).\nonumber
\end{eqnarray}
Comparing the coefficients, one obtains $x=y=z=0$, $n=\alpha+\pi$, and
\begin{eqnarray}
{m}=\arccos{{(|V_{td}||V_{tb}|)^2-[(|V_{ud}||V_{ub}|)^2+(|V_{cd}||V_{cb}|)^2]}\over{2|V_{ud}||V_{ub}||V_{cd}||V_{cb}|}}\;.
\end{eqnarray}
Therefore, the transformation from case $\alpha_1$ to case $\alpha_2$ is achieved by,
\begin{eqnarray}
{V^{\alpha_{2}}_{CKM}}=
\left(
\begin{array}{ccc}
1 & 0 &0           \\
0 &e^{im} &0 \\
0&0 &-e^{i\alpha}
\end{array}
\right)
{V^{\alpha_{1}}_{CKM}}\;.
\end{eqnarray}

Similarly one can transform the other two cases to case $\alpha_2$
too.

We now discuss the $\beta$ parametrization. From eq.\ref{phases}, we can have four choices for the location of the
phase similar to the $\alpha$ parametrization.  They are
\begin{eqnarray}
&&\beta_1)\;. \;\;(|V_{cd}|, |V_{cb}|, |V_{td}|, -|V_{tb}|e^{i\beta})\;,\nonumber\\
&&\beta_2)\;. \;\;(|V_{cd}|, |V_{cb}|, -|V_{td}|e^{-i\beta}, |V_{tb}|)\;,\nonumber\\
&&\beta_3)\;. \;\;(|V_{cd}|, -|V_{cb}|e^{-i\beta}, |V_{td}|, V_{tb}|)\;,\nonumber\\
&&\beta_4)\;. \;\;(- |V_{cd}|e^{i\beta}, |V_{cb}|, |V_{td}|, |V_{tb}|)\;.
\end{eqnarray}

In a similar way as for the four cases of $\alpha_i$, one can show that the above four cases are equivalent. Detailed expressions for these four cases are given in the Appendix. We will display the explicit form
for case $\beta_2$. We have
\begin{equation}
V^{\beta_2}_{CKM} = \left(
\begin{array}{ccc}
\frac{|V_{td}||V_{tb}|e^{-i\beta}-|V_{cb}||V_{cd}|}{|V_{ub}|}& \frac{(|V_{cd}|^{2}-|V_{ub}|^{2})|V_{cb}|-|V_{cd}||V_{td}||V_{tb}|e^{-i\beta}}{|V_{cs}||V_{ub}|} &|V_{ub}|\\
|V_{cd}|&|V_{cs}|
 &|V_{cb}|\\
 -|V_{td}|e^{-i\beta} & \frac{|V_{cd}||V_{td}|e^{-i\beta}-|V_{cb}||V_{tb}|}{|V_{cs}|} & |V_{tb}|\\
\end{array}
\right)\;.
\end{equation}
For this case it is convenient to take $\beta$,~$|V_{cs}|$,~$|V_{cb}|$,~$|V_{tb}|$ as independent variables. The other quantities can be expressed as
\begin{eqnarray*}
|V_{cd}|&=&\sqrt{1-|V_{cs}|^{2}-|V_{cb}|^{2}},|V_{ub}|=\sqrt{1-|V_{cb}|^{2}-|V_{tb}|^{2}},\\
|V_{td}|&=&\frac{|V_{tb}||V_{cd}||V_{cb}|\cos{\beta}}{1-|V_{cb}|^{2}}\\
&+&\sqrt{(\frac{|V_{tb}||V_{cd}||V_{cb}|\cos{\beta}}{1-|V_{cb}|^{2}})^{2}-\frac{|V_{ub}|^{2}(|V_{cd}|^{2}-1)+|V_{cd}|^{2}|V_{cb}|^{2}}{1-|V_{cb}|^{2}}}\;.
\end{eqnarray*}
The Jarlskog parameter $J$ is given by
\begin{eqnarray*}
J=|V_{cb}||V_{tb}||V_{cd}||V_{td}|\sin{\beta}.
\end{eqnarray*}

For $\gamma$ parametrization, the four different cases are defined by
\begin{eqnarray}
&&\gamma_1)\;. \;\;(|V_{ud}|, |V_{ub}|, |V_{cd}|, -|V_{cb}|e^{i\gamma})\;,\nonumber\\
&&\gamma_2)\;. \;\;(|V_{ud}|, |V_{ub}|, -|V_{cd}|e^{-i\gamma}, |V_{cb}|)\;,\nonumber\\
&&\gamma_3)\;. \;\;(|V_{ud}|, -|V_{ub}|e^{-i\gamma}, |V_{cd}|, V_{cb}|)\;,\nonumber\\
&&\gamma_4)\;. \;\;(- |V_{ud}|e^{i\gamma}, |V_{ub}|, |V_{cd}|, |V_{cb}|)\;.
\end{eqnarray}

The detailed expressions for these four cases are given in Appendix.
We display the explicit form for case $\gamma_3$ here. We have
\begin{equation}
V^{\gamma_3}_{CKM} = \left(
\begin{array}{ccc}
|V_{ud}|& -\frac{|V_{ud}||V_{cd}|-|V_{ub}||V_{cb}|e^{-i\gamma}}{|V_{cs}|} &-|V_{ub}|e^{-i\gamma}\\
|V_{cd}|&|V_{cs}|
 &|V_{cb}|\\
|V_{td}| & \frac{(|V_{cb}|^{2}-|V_{td}|^{2})|V_{cd}|-|V_{cb}||V_{ud}||V_{ub}|e^{-i\gamma}}{|V_{cs}||V_{td}|} &  \frac{|V_{ud}||V_{ub}|e^{-i\gamma}-|V_{cd}||V_{cb}|}{|V_{td}|}\\
\end{array}
\right)\;.
\end{equation}
Taking ~$\gamma$,~$|V_{cd}|$,~$|V_{cs}|$,~$|V_{td}|$~as variables, the other quantities can be expressed as
\begin{eqnarray*}
|V_{ud}|&=&\sqrt{1-|V_{cd}|^{2}-|V_{td}|^{2}},|V_{cb}|=\sqrt{1-|V_{cd}|^{2}-|V_{cs}|^{2}},\\
|V_{ub}|&=&\frac{|V_{ud}||V_{cd}||V_{cb}|\cos{\gamma}}{1-|V_{cd}|^{2}}\\
&-&\sqrt{(\frac{|V_{ud}||V_{cd}||V_{cb}|\cos{\gamma}}{1-|V_{cd}|^{2}})^{2}-\frac{|V_{cs}|^{2}(|V_{ud}|^{2}-1)+|V_{ud}|^{2}|V_{cd}|^{2}}{1-|V_{cd}|^{2}}},
\end{eqnarray*}
and
\begin{eqnarray*}
J=|V_{ub}||V_{cb}||V_{ud}||V_{cd}|\sin{\gamma}.
\end{eqnarray*}

\section{Relations between different parameterizations}

We now show that the $\alpha$, $\beta$ and $\gamma$ parameterizations can
also be transformed from each other. The twelve parameterizations discussed before in the previous section are all equivalent.
We have already shown that the four parameterizations for $\alpha$ or $\beta$ or $\gamma$ are equivalent ones,
we will therefore only need to show that one of the $\alpha$ parameterizations is equivalent to one of the $\beta$ or $\gamma$ parameterizations.

We now show that $V_{CKM}^{\alpha_3}$ and $V_{CKM}^{\beta_3}$. For these two parametrizations, the first column and the third row are already identical in these two parameterizations.
Using
\begin{equation}
|V_{ub}||V_{ud}|e^{i\alpha}+|V_{cb}||V_{cd}|e^{-i\beta}=|V_{td}||V_{tb}|.
\end{equation}
One can readily show that the 12, 13, 22 and 23 entries of $V^{\alpha_3}_{CKM}$ and $V_{CKM}^{\beta_3}$ are equal. Therefore
\begin{eqnarray}
V^{\alpha_3}_{CKM}& = &\left(
\begin{array}{ccc}
|V_{ud}|&- \frac{|V_{ud}||V_{td}|-|V_{tb}||V_{ub}|e^{i\alpha}}{|V_{ts}|} & - |V_{ub}|e^{i\alpha}\\
|V_{cd}|&
 \frac{(|V_{tb}|^{2}-|V_{cd}|^{2})|V_{td}|-|V_{ub}||V_{ud}||V_{tb}|e^{i\alpha}}{|V_{ts}||V_{cd}|}&\frac{|V_{ud}||V_{ub}|e^{i\alpha}-|V_{td}||V_{tb}|}{|V_{cd}|}\\
 |V_{td}| & |V_{ts}|& |V_{tb}|\\
\end{array}
\right)\\
&=&\left(
\begin{array}{ccc}
|V_{ud}|& -\frac{(|V_{ud}|^{2}-|V_{tb}|^{2})|V_{td}|+|V_{tb}||V_{cd}||V_{cb}|e^{-i\beta}}{|V_{ts}||V_{ud}|} & -\frac{|V_{td}||V_{tb}|-|V_{cb}||V_{cd}|e^{-i\beta}}{|V_{ud}|}\\
|V_{cd}|&
 \frac{|V_{cb}||V_{tb}|e^{-i\beta}-|V_{cd}||V_{td}|}{|V_{ts}|}&-|V_{cb}|e^{-i\beta}\\
 |V_{td}| & |V_{ts}| & |V_{tb}|\\
\end{array}
\right) = V^{\beta_3}_{CKM}.\nonumber
\end{eqnarray}

Similarly, using
\begin{eqnarray}
|V_{tb}||V_{td}|e^{i\alpha}+|V_{cb}||V_{cd}|e^{-i\gamma}=|V_{ud}||V_{ub}|,\\
|V_{ud}||V_{ub}|e^{-i\gamma}+|V_{tb}||V_{td}|e^{i\beta}=|V_{cd}||V_{cb}|,
\end{eqnarray}
we obtain
\begin{equation}
V_{CKM}^{\alpha2}=V_{CKM}^{\gamma2},V_{CKM}^{\beta1}=V_{CKM}^{\gamma3}.
\end{equation}

We therefore have shown explicitly that  the $\alpha$, $\beta$ and
$\gamma$ parameterizations are related and can be transformed from
one to another.

Numerically, one finds that the approximate relations $\delta_{KM} \approx \alpha$ and $\delta_{PDG}\approx \gamma$.
These can be understood easily by noticing the relations between them~\cite{fh,koide},
\begin{eqnarray}
&&\alpha
=\arctan({\sin \delta_{KM} \over x_{\alpha}-\cos\delta_{KM} }),\;\;\;\;x_{\alpha} = {c_1s_2s_3\over c_2c_3} = {|V_{ud}||V_{td}||V_{ub}|\over |V_{cd}||V_{us}|}=0.0006.\nonumber\\
&&\gamma=\arctan({\sin \delta_{PDG} \over x_{\gamma}+\cos\delta_{PDG} }),\;\;\;\;x_{\gamma}= {c_{12}s_{23}s_{13}\over s_{12} c_{23}} = {|V_{ud}||V_{cb}||V_{ub}|\over |V_{tb}||V_{us}|} =0.0006.\nonumber
\end{eqnarray}
Therefore, $\delta_{KM} + \alpha$ is approximately $\pi$, since $\alpha$ is close to $90^\circ$, $\delta_{KM}$ must also be close to $90^\circ$ and therefore $\delta_{KM} \approx \alpha$. It is also clear that $\delta_{PDG}$ is approximately equal to $\gamma$.

One may wonder if there is a parametrization where the phase is
close to $\beta$. We find indeed there are angle prametrizations in
which the CP violating phase is close to $\beta$. An example is
provided by the parametrization $P4$ discussed in Ref.~\cite{xing}
where
\begin{equation}
 {V_{CKM}^{P4}}
 = \left(
\begin{array}{ccc}
c_{\theta}c_{\tau}& c_{\theta}s_{\sigma}s_{\tau}+ s_{\theta}c_{\sigma}e^{-i\varphi} & c_{\theta}c_{\sigma}s_{\tau}-s_{\theta}s_{\sigma}e^{-i\varphi}\\
-s_{\theta} c_{\tau}&-s_\theta s_\sigma s_\tau + c_\theta c_\sigma e^{-i\varphi}
 &-s_{\theta} c_{\sigma} s_\tau - c_\theta s_\sigma e^{-i\varphi}\\
 -s_{\tau} & s_{\sigma}c_{\tau} &
 c_{\sigma}c_{\tau}\\
\end{array}
\right).
\end{equation}
We have
\begin{eqnarray}
&&\beta
=\arctan({\sin \varphi \over x_{\beta}+ \cos\varphi }),\;\;\;\;x_{\beta} = {s_\theta c_\sigma s_\tau\over c_\theta s_\sigma} = {|V_{cd}||V_{tb}||V_{td}|\over |V_{ud}||V_{ts}|} = 0.0497.
\end{eqnarray}

One may also wonder if one can find a parametrization in which the
CP violating angle is one of the $\alpha$, $\beta$ and $\gamma$, and
the other three quantities to parameterize the mixing matrix can be
chosen to be three angles similar to those in the PDG or KM
paramerizations. We find that this is impossible. It has been shown
that there are nine different ways to parameterize the mixing matrix
using one phase and three angles~\cite{xing}. Explicit inspections
find that none of the phases can be exactly, allowing sign
differences or plus or minus a $\pi$, identified as one of the
$\alpha$, $\beta$ or $\gamma$. The reason is that in the $\alpha$,
$\beta$ and $\gamma$ parametrization we have introduced, one needs
one of the elements in mixing matrix to be of the form
$|V_{ij}|e^{\pm i\alpha}$, $|V_{ij}|e^{\pm i\beta}$, or
$|V_{ij}|e^{\pm i\gamma}$, and also another five real matrix
elements which cannot be satisfied with only three angles.

\section{Wolfenstein-like Expansions}

It has proven to be convenient to use approximate formula such as the Wolfenstein parametrization\cite{wolfenstein}.
In the literature different approximate forms have been proposed\cite{he-ma,ma-wolf}.
In this section, we discuss the Wolfenstein-like expansions in the $\alpha$, $\beta$ and $\gamma$ parameterizations demanding
to use one of the $\alpha$, $\beta$ and $\gamma$ as one of the parameters.

For the $\alpha_i)$ cases, we find it convenient to work with $\alpha_1)$ case. We use
$|V_{us}| = \lambda$, $|V_{ub}| = a \lambda^3$, $|V_{td}| = b\lambda^3$ and $\alpha$ as parameters.
The numerical values of $\lambda$, $a$ and $b$ are determined to be
$\lambda= 0.2252\pm 0.0010$, $a= 0.3170\pm0.0130 $, and
$b= 0.7670\pm0.0250 $.
To order
$\lambda^3$, we have
\begin{equation}
V_{CKM}^{\alpha_{1}}\approx  \left(
\begin{array}{ccc}
{1-{1\over 2}{\lambda}^2}& \lambda & a{\lambda}^3\\
\lambda&{-1+{1\over 2}{\lambda}^2} &-(a-be^{-i\alpha})\lambda^2\\
b{\lambda}^3 & (a{e}^{-i\alpha}-b){\lambda}^2 & -e^{-i\alpha}\\
\end{array}
\right).
\end{equation}

One can further rotate the phase of c-quark by $\pi$ and b-quark by $\pi +\alpha$ to obtain an expression where the diagonal
entries are close to 1. We obtain
\begin{equation}
V_{CKM}^{\alpha_{1}}\approx \left(
\begin{array}{ccc}
{1-{1\over 2}{\lambda}^2}& \lambda & - a{\lambda}^3 e^{i\alpha}\\
-\lambda&{1-{1\over 2}{\lambda}^2} &-(ae^{i\alpha}-b)\lambda^2\\
b{\lambda}^3 & (a{e}^{-i\alpha}-b){\lambda}^2 & 1\\
\end{array}
\right).
\end{equation}
The above expansion is equivalent to that discussed in
Ref.\cite{ma-wolf}. The parameters $\delta$, $h$ and $f$ in
Ref.\cite{ma-wolf} are related to the above parameters by, $\delta =
-\alpha$, $f = b$ and $h = -a$. At more than $\lambda^3$ order,
there are differences for our approximation and that proposed in
Ref. \cite{ma-wolf}.

For $\beta_i)$ cases, it is convenient to use $\beta_1)$
for expansion. Setting $|V_{cd}|= \lambda$,
$|V_{td}| = b \lambda^3$, and  $|V_{cb}| = c \lambda^2$ with $\lambda=
0.2251\pm 0.0010$, and  $b=0.7685\pm0.0250 $
$c= 0.8185\pm0.0176$. Rotating
the b-quark field by a phase $\pi -\beta$, we obtain to order
$\lambda^3$
\begin{equation}
V_{CKM}^{\beta_{1}} \approx \left(
\begin{array}{ccc}
{1-{1\over 2}{\lambda}^2}& -\lambda & {\lambda}^3(c e^{-i\beta}-b)\\
\lambda & 1-{1\over 2}{\lambda}^2 & -c\lambda^2 e^{-i\beta}\\
b{\lambda}^3 & c\lambda^2{e}^{i\beta} & 1\\
\end{array}
\right).
\end{equation}

For the $\gamma_i)$ cases, $\gamma_4)$ case is convenient for expansion. Setting  $|V_{cd}|= \lambda$, $|V_{ub}|={a{\lambda}^3}$  and $|V_{cb}|={c{\lambda}^2}$
with $\lambda= 0.2251\pm0.0010$, $a= 0.3176\pm0.0130$, and $c= 0.8185\pm0.0176$.
Rotating d-quark by a phase $\pi$ and u-quark by a phase $-\gamma$, we obtain to order $\lambda^3$
\begin{equation}
V_{CKM}^{\gamma_4} \approx \left(
\begin{array}{ccc}
1-{1\over 2}{\lambda}^2& \lambda & a{\lambda}^3 e^{- i\gamma}\\
- \lambda & 1-{1\over 2}{\lambda}^2 & c\lambda^2\\
- {\lambda}^3(ae^{i\gamma}-c) & - c\lambda^2 & 1\\
\end{array}
\right).
\end{equation}

To $\lambda^3$ order, the above expansion is equivalent to the Wolfenstein parametrization\cite{wolfenstein}.
The parameters $A$, $\rho$ and $\eta$ in the Wolfenstein parametrization are related to $a$, $c$ and $\gamma$ by
$c= A$, $\rho = a\cos\gamma/c$ and $\eta = a\sin\gamma/c$. Again at more than $\lambda^3$ order, there are differences in these two approximations.

Among the three phase angles, $\alpha$, $\beta$ and $\gamma$, the
best measured one is $\beta$. This makes the approximate expression
$V^{\beta_1}_{CKM}$ better than others.

\section{Conclusion}

To conclude, we have proposed new parametrizations of the CKM matrix
using the three measurable phase angles $\alpha$, $\beta$ and
$\gamma$ in the unitarity triangle as the CP violating phase. For
each of the  $\alpha$, $\beta$ and $\gamma$ parametrization there
are four ways to parameterize the mixing matrix where one column and
one row elements are all real. We have shown explicitly that all
these cases are equivalent. We have studied relations of the
$\alpha$, $\beta$ and $\gamma$ parametrizations with the usual three
rotation angles in flavor space and one CP violating phase
parametrizations. We find that it is not possible to parameterize
the CKM matrix using three angles and taking one of the $\alpha$,
$\beta$ and $\gamma$ as the CP violating phase. However, there are
rotation angle parametrizations in which the CP violating are very
close to one of the phase angles $\alpha$, $\beta$ or $\gamma$. We,
however, emphasis that the $\alpha$, $\beta$ and $\gamma$
parametrizations we proposed have the advantage that all the
elements used to describe the mixing matrix are physically
measurable quantities, unlike the parametrizations of using rotation
angles in flavor space and a phase whose values are parametrization
dependent. The $\alpha$, $\beta$ and $\gamma$ parametrizations are
parametrization independent representation of the CKM matrix. We
also suggest new Wolfenstein-like paramterizations.
\\

\noindent
{\bf Acknowledgement}
This work was supported in part by NSC and NCTS of ROC, NNSF(grant No:11175115) and Shanghai science and technology commission (grant no: 11DZ2260700) of PRC.

\appendix
\section{Expressions of $V_{CKM}$ for $\beta_i$ and $\gamma_i$ cases}

Expressions for the four $\beta$ and four $\gamma$ parameterizations.

\begin{equation}
V_{CKM}^{\beta_1} = \left(
\begin{array}{ccc}
|V_{ud}|& -{\frac{(|V_{ud}|^{2}-|V_{cb}|^{2})|V_{cd}|+|V_{cb}||V_{td}||V_{tb}|e^{i\beta}}{|V_{cs}||V_{ud}|}} &-{\frac{|V_{cb}||V_{cd}|-|V_{td}||V_{tb}|e^{i\beta}}{|V_{ud}|}}\\
|V_{cd}|&|V_{cs}| &|V_{cb}|\\
|V_{td}| & \frac{|V_{cb}||V_{tb}|e^{i\beta}-|V_{cd}||V_{td}|}{|V_{cs}|} & -|V_{tb}|e^{i\beta}\\
\end{array}
\right)
\end{equation}\\

\begin{equation}
V_{CKM}^{\beta_2} = \left(
\begin{array}{ccc}
\frac{|V_{td}||V_{tb}|e^{-i\beta}-|V_{cb}||V_{cd}|}{|V_{ub}|}& \frac{(|V_{cd}|^{2}-|V_{ub}|^{2})|V_{cb}|-|V_{cd}||V_{td}||V_{tb}|e^{-i\beta}}{|V_{cs}||V_{ub}|} &|V_{ub}|\\
|V_{cd}|&|V_{cs}|
 &|V_{cb}|\\
 -|V_{td}|e^{-i\beta} & \frac{|V_{cd}||V_{td}|e^{-i\beta}-|V_{cb}||V_{tb}|}{|V_{cs}|} & |V_{tb}|\\
\end{array}
\right)
\end{equation}\\

\begin{equation}
V_{CKM}^{\beta_3} = \left(
\begin{array}{ccc}
|V_{ud}|& -\frac{(|V_{ud}|^{2}-|V_{tb}|^{2})|V_{td}|+|V_{tb}||V_{cd}||V_{cb}|e^{-i\beta}}{|V_{ts}||V_{ud}|} & -\frac{|V_{td}||V_{tb}|-|V_{cb}||V_{cd}|e^{-i\beta}}{|V_{ud}|}\\
|V_{cd}|&\frac{|V_{cb}||V_{tb}|e^{-i\beta}-|V_{cd}||V_{td}|}{|V_{ts}|}&-|V_{cb}|e^{-i\beta}\\
|V_{td}| & |V_{ts}| & |V_{tb}|\\
\end{array}
\right)
\end{equation}\\

\begin{equation}
V_{CKM}^{\beta_4} = \left(
\begin{array}{ccc}
-\frac{|V_{td}||V_{tb}|-|V_{cb}||V_{cd}|e^{i\beta}}{|V_{ub}|}& -\frac{(|V_{ub}|^{2}-|V_{td}|^{2})|V_{tb}|+|V_{td}||V_{cb}||V_{cd}|e^{i\beta}}{|V_{ts}||V_{ub}|} & |V_{ub}|\\
-|V_{cd}|e^{i\beta}&-\frac{|V_{cb}||V_{tb}|-|V_{cd}||V_{td}|e^{i\beta}}{|V_{ts}|}&|V_{cb}|\\
|V_{td}| & |V_{ts}| & |V_{tb}|\\
\end{array}
\right)
\end{equation}\\

\begin{equation}
V_{CKM}^{\gamma_1} = \left(
\begin{array}{ccc}
|V_{ud}|& |V_{us}| &|V_{ub}|\\
|V_{cd}|&-\frac{|V_{ud}||V_{cd}|-|V_{ub}||V_{cb}|e^{i\gamma}}{|V_{us}|}
 &-|V_{cb}|e^{i\gamma}\\
|V_{td}| & \frac{(|V_{ub}|^{2}-|V_{td}|^{2})|V_{ud}|-|V_{ub}||V_{cd}||V_{cb}|e^{i\gamma}}{|V_{us}||V_{td}|} & -\frac{|V_{ud}||V_{ub}|-|V_{cd}||V_{cb}|e^{i\gamma}}{|V_{td}|}\\
\end{array}
\right)
\end{equation}\\

\begin{equation}
V_{CKM}^{\gamma_2} = \left(
\begin{array}{ccc}
|V_{ud}|& |V_{us}| &|V_{ub}|\\
-|V_{cd}|e^{-i\gamma}&\frac{|V_{ud}||V_{cd}|e^{-i\gamma}-|V_{ub}||V_{cb}|}{|V_{us}|}
 &|V_{cb}|\\
 \frac{|V_{cd}||V_{cb}|e^{-i\gamma}-|V_{ud}||V_{ub}|}{|V_{tb}|} & \frac{(|V_{ud}|^{2}-|V_{tb}|^{2})|V_{ub}|-|V_{ud}||V_{cd}||V_{cb}|e^{-i\gamma}}{|V_{us}||V_{tb}|} &|V_{tb}| \\
\end{array}
\right)
\end{equation}\\

\begin{equation}
V_{CKM}^{\gamma_3} = \left(
\begin{array}{ccc}
|V_{ud}|& -\frac{|V_{ud}||V_{cd}|-|V_{ub}||V_{cb}|e^{-i\gamma}}{|V_{cs}|} &-|V_{ub}|e^{-i\gamma}\\
|V_{cd}|&|V_{cs}| &|V_{cb}|\\
|V_{td}| & \frac{(|V_{cb}|^{2}-|V_{td}|^{2})|V_{cd}|-|V_{cb}||V_{ud}||V_{ub}|e^{-i\gamma}}{|V_{cs}||V_{td}|} &  \frac{|V_{ud}||V_{ub}|e^{-i\gamma}-|V_{cd}||V_{cb}|}{|V_{td}|}\\
\end{array}
\right)
\end{equation}\\

\begin{equation}
V_{CKM}^{\gamma_4} = \left(
\begin{array}{ccc}
-|V_{ud}|e^{i\gamma}& \frac{|V_{ud}||V_{cd}|e^{i\gamma}-|V_{ub}||V_{cb}|}{|V_{cs}|} &|V_{ub}|\\
|V_{cd}|&|V_{cs}|
 &|V_{cb}|\\
 \frac{|V_{ud}||V_{ub}|e^{i\gamma}-|V_{cd}||V_{cb}|}{|V_{tb}|} & \frac{(-|V_{tb}|^{2}+|V_{cd}|^{2})|V_{cb}|-|V_{cd}||V_{ud}||V_{ub}|e^{i\gamma}}{|V_{cs}||V_{tb}|} & |V_{tb}|\\
\end{array}
\right)
\end{equation}

\end{document}